\documentclass[sigconf]{acmart}

\usepackage{balance}
\usepackage{subcaption}
\usepackage{multirow}
\usepackage{graphicx}
\usepackage{float}
\usepackage{amsfonts,amssymb}

\def\BibTeX{{\rm B\kern-.05em{\sc i\kern-.025em b}\kern-.08emT\kern-.1667em\lower.7ex\hbox{E}\kern-.125emX}}
    
\copyrightyear{2019} 
\acmYear{2019} 
\acmConference[MM '19]{Proceedings of the 27th ACM International Conference on Multimedia}{October 21--25, 2019}{Nice, France}
\acmBooktitle{Proceedings of the 27th ACM International Conference on Multimedia (MM '19), October 21--25, 2019, Nice, France}
\acmPrice{15.00}
\acmDOI{10.1145/3343031.3351022}
\acmISBN{978-1-4503-6889-6/19/10}

\setcopyright{acmcopyright}
\acmSubmissionID{fp749}

\begin{document}

\fancyhead{}

\title{Progressive Image Inpainting with Full-Resolution Residual Network}

\author{Zongyu Guo}
\affiliation{%
  \institution{University of Science and Technology of China}
  \city{Hefei}
  \country{China}
}
\email{guozy@mail.ustc.edu.cn}

\author{Zhibo Chen}
\authornote{Corresponding author.}
\affiliation{%
  \institution{University of Science and Technology of China}
  \city{Hefei}
  \country{China}
}
\email{chenzhibo@ustc.edu.cn}

\author{Tao Yu}
\affiliation{%
  \institution{University of Science and Technology of China}
  \city{Hefei}
  \country{China}
}
\email{yutao666@mail.ustc.edu.cn}

\author{Jiale Chen}
\affiliation{%
  \institution{University of Science and Technology of China}
  \city{Hefei}
  \country{China}
}
\email{chenjlcv@mail.ustc.edu.cn}
 
\author{Sen Liu}
\affiliation{%
  \institution{University of Science and Technology of China}
  \city{Hefei}
  \country{China}
}
\email{elsen@iat.ustc.edu.cn}

%
\renewcommand{\shortauthors}{Guo, et al.}

%
\begin{abstract}
Recently, learning-based algorithms for image inpainting achieve remarkable progress dealing with squared or irregular holes. However, they fail to generate plausible textures inside damaged area because there lacks surrounding information. A progressive inpainting approach would be advantageous for eliminating central blurriness, \textit{i.e.}, restoring well and then updating masks. In this paper, we propose full-resolution residual network (FRRN) to fill irregular holes, which is proved to be effective for progressive image inpainting. We show that well-designed residual architecture facilitates feature integration and texture prediction. Additionally, to guarantee completion quality during progressive inpainting, we adopt \textit{N Blocks, One Dilation} strategy, which assigns several residual blocks for one dilation step. Correspondingly, a step loss function is applied to improve the performance of intermediate restorations. The experimental results demonstrate that the proposed FRRN framework for image inpainting is much better than previous methods both quantitatively and qualitatively. Our codes are released at: \url{https://github.com/ZongyuGuo/Inpainting_FRRN}.
\end{abstract}

%
%

\ccsdesc[500]{Applied computing~Fine arts}

%
\keywords{progressive image inpainting; irregular mask; residual network; computer vision}

%
\begin{teaserfigure}
\centering
    \begin{subfigure}{0.19\textwidth}
     \includegraphics[width=\textwidth]{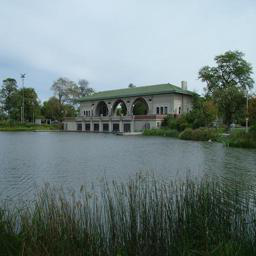}
     \caption{Ground Truth}
    \end{subfigure}
    \hspace{0.0015\textwidth}
    \begin{subfigure}{0.19\textwidth}
     \includegraphics[width=\textwidth]{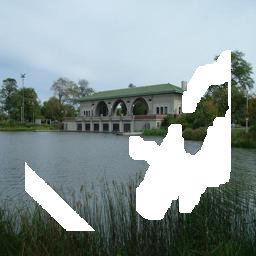}
     \caption{Damaged image}
    \end{subfigure}
    \hspace{0.0015\textwidth}
    \begin{subfigure}{0.19\textwidth}
     \includegraphics[width=\textwidth]{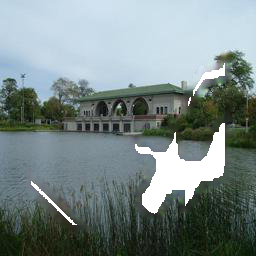}
     \caption{Once dilation}
    \end{subfigure}
    \hspace{0.0015\textwidth}
    \begin{subfigure}{0.19\textwidth}
     \includegraphics[width=\textwidth]{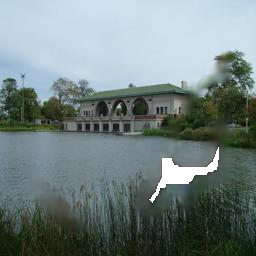}
     \caption{Twice dilation}
    \end{subfigure}
    \hspace{0.0015\textwidth}
    \begin{subfigure}{0.19\textwidth}
     \includegraphics[width=\textwidth]{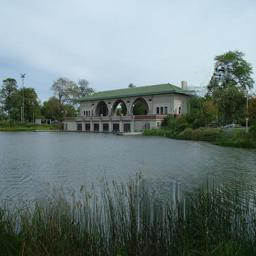}
     \caption{Final recovering}
    \end{subfigure}
    \caption{Progressive image inpainting for irregular holes.}
  \label{fig:teaser}
\end{teaserfigure}

%
\maketitle

\section{Introduction}

Image inpainting targets at reconstructing missing areas in corrupted images which can be synthesized by natural images and binary masks. This technique is indispensable in many interactive photography applications, such as photo dis-occlusion, object removal, error concealment and so on. Traditional diffusion-based methods \cite{ballester2000filling,esedoglu2002digital,bertalmio2003simultaneous,levin2003learning,liu2007image} are able to utilize texture or structure information in known areas to reconstruct missing areas. These diffusion-based approaches can be considered as progressive processes to restore damaged images step-by-step. However, these approaches will cause error propagation problem because failed restoration at one inpainting step will generate worse results at the following steps. 

Over the past few years, Deep Neural Networks (DNNs) have shown prominent capability for learning data priors. Most existing learning-based methods for image inpainting \cite{pathak2016context,yeh2017semantic,isola2017image,yu2018generative,DBLP:journals/corr/abs-1903-04227,xiong2019foreground,nazeri2019edgeconnect} do not consider progressive inpainting strategy, which is a more natural way to reconstruct missing areas. Designing a reasonable progressive inpainting network, which can better utilize prior information of data distribution, is still challenging. Zhang et al. \cite{zhang2018semantic} adopt four-step progressive generative networks (PGNs) to complete corrupted images with squared masks and achieves promising results. However, they cannot deal with irregular masks or squared masks of various sizes, which are common in real applications such as arbitrary object removal in photos. It is hard but more practical to build a progressive architecture which can adaptively fill irregular holes and simultaneously assure restoration quality during the whole completion process.

To cooperate with progressive inpainting for irregular holes, a residual architecture is suggested by us to correctly reconstruct interior regions inside holes. In this paper, we propose full-resolution residual block (FRRB) which can generate residuals and update masks (erode holes) if required (Figure \ref{FRRB_arch}). Compared with PConv \cite{liu2018image}, which can shrink masks after every partial convolution layer, proposed block-level inpainting architecture can utilize dilation module to complete images more effectively and accurately. New generated residuals will be added to input of current FRRB and turn to be fed into next FRRB. As for its inner structure, we maintain a reliable full-resolution branch parallel with a feature integration branch. This full-resolution branch can facilitate texture prediction and accelerate convergence, which is proved by an ablation study. 

Based on designed FRRB, we propose an architecture of full-resolution residual network (FRRN) for progressive image inpainting. High quality of intermediate restorations is required to inhibit error propagation. We adopt \textit{N Blocks, One Dilation} strategy to preserve restoration quality, \textit{i.e.}, restoring well and then updating masks. Assigning more FRRBs in one mask-update period would enhance the capability of dilation module. All dilation modules can jointly reconstruct corrupted images step-by-step (Figure \ref{fig:teaser}). Further, we adopt step loss function, which is imposed to improve middle restoration quality. Experimental results show that both \textit{N Blocks, One Dilation} strategy and step loss function contribute to the state-of-the-art performance of our method.  

Compared with previous algorithms, proposed FRRN performs much better both quantitatively and qualitatively when evaluated on Places2 \cite{zhou2017places}. We also conduct experiments on CelebA \cite{liu2015deep} and show its promising performance. 

In short, our main contributions can be summarized as follows:
\begin{itemize}
    \item We design an effective full-resolution residual block (FRRB), which acts on missing areas and is able to dilate clean regions. Its inner architecture which maintains a full-resolution branch facilitates feature integration and texture prediction.
    \item We propose full-resolution residual network (FRRN) and achieve the state-of-the-art performance for image inpainting. Proposed FRRN which contains several dilation modules can complete irregular holes progressively.
    \item We adopt a novel \textit{N Blocks, One Dilation} strategy to constrain the quality of middle reconstructions and correspondingly apply step loss function. We demonstrate that both \textit{N Blocks, One Dilation} strategy and step loss function contribute to final promising completion results.
\end{itemize}

\section{Related work}

Existing work on image inpainting can be commonly divided into two categories: traditional methods and learning-based methods. In this section, we briefly overview these two ways for image inpainting and then summarize the effectiveness of residual architecture in DNNs. 

\subsection{Traditional Methods}

When it comes to traditional inpainting methods, we can further divide them into two sub-categories: diffusion-based methods \cite{bertalmio2000image,ballester2000filling,esedoglu2002digital,bertalmio2003simultaneous,levin2003learning,liu2007image} and patch-based methods \cite{drori2003fragment,sun2005image,barnes2009patchmatch,xu2010image,darabi2012image,huang2014image}. Diffusion-based methods usually fill missing regions with information surrounding them, \textit{i.e.}, propagate neighboring information into missing regions. However, they fail to synthesize meaningful content since the information only comes from its neighborhood and thus cannot solve the case of large missing regions. Patch-based methods attempt to search regions that are the most similar or relevant to missing regions in a same image and then copy them to corresponding missing regions. These methods often aim to minimize discontinuities in an iterative way \cite{darabi2012image,huang2014image}. These methods are able to synthesize relatively smooth and satisfactory results but spend a high computational cost. To address the computation issue, a randomized algorithm for quickly finding approximate nearest neighbor matches between image patches is proposed in PatchMatch \cite{barnes2009patchmatch} which greatly speeds up searching and also provides quite good results. These traditional methods fail to generate detailed structures since they only process on low-level local appearance and cannot capture high-level semantic information.

\subsection{Learning-based Methods}

Learning-based methods have shown promising inpainting results thanks to the rapid development of deep neural networks (DNNs) and generative adversarial networks (GANs) \cite{goodfellow2014generative}. These methods are able to improve inpainting results by using sematic information. One of pioneer works Context Encoder \cite{pathak2016context} uses an encoder-decoder architecture to fill missing regions. Further, GAN-based works \cite{yeh2017semantic,isola2017image,yu2018generative,zhang2018semantic,DBLP:journals/corr/abs-1903-04227,xiong2019foreground,nazeri2019edgeconnect} can generate visually plausible results. \cite{Vo:2018:SI:3240508.3240678} adopts structural loss to preserve reconstruction of edge. Liu et al. \cite{liu2018image} use partial convolutions where the convolution is only conditioned on valid pixels to reduce artifacts caused by distribution difference between masked area and non-masked area. Most existing learning-based methods \cite{pathak2016context,yeh2017semantic,isola2017image,yu2018generative,DBLP:journals/corr/abs-1903-04227,xiong2019foreground,nazeri2019edgeconnect,Vo:2018:SI:3240508.3240678} do not consider progressive inpainting policy which is a more natural and comprehensive way to complete missing regions except \cite{zhang2018semantic}. Zhang et al. \cite{zhang2018semantic} divide the hole filling process into several phases to progressively shrink large missing regions and yield promising inpainting results. However, they target at filling holes with the same shape, \textit{i.e.}, squared or oval holes. They cannot deal with irregular holes or regular masks with various sizes in one network. 

\subsection{Residual Architecture}

ResNet \cite{he2016deep}, proposed by He et al., has strongly impacted the development of DNNs thanks to its effective structure, which can deepen network but avoid vanishing gradient problem at training stage \cite{he2016deep}. It can be applied into many computer vision tasks, such as denoising \cite{zhang2017beyond} and super-resolution \cite{kim2016accurate,lim2017enhanced}. In these practical applications, residual architecture benefits not only training process for network convergence, but also these tasks themselves. When utilized in these tasks, every residual block only requires to recover tiny difference value for final reconstruction. A residual architecture for progressive image inpainting may also be advantageous because every residual block only targets at reconstructing specified region and stacking residual blocks is exactly suitable for this progressive inpainting strategy. 

\section{Approach}

We propose a full-resolution residual network (FRRN) for progressive image inpainting with irregular holes. In this section, we start with analysis of designed full-resolution residual block (FRRB), a component of FRRN. Then we introduce the inner architecture of FRRN and illustrate \textit{N Blocks, One Dilation} strategy for network building. Finally, we discuss loss functions including middle step loss.

\begin{figure}[t]
\includegraphics[scale=0.85, clip, trim=10.8cm 3.8cm 8.5cm  2cm]{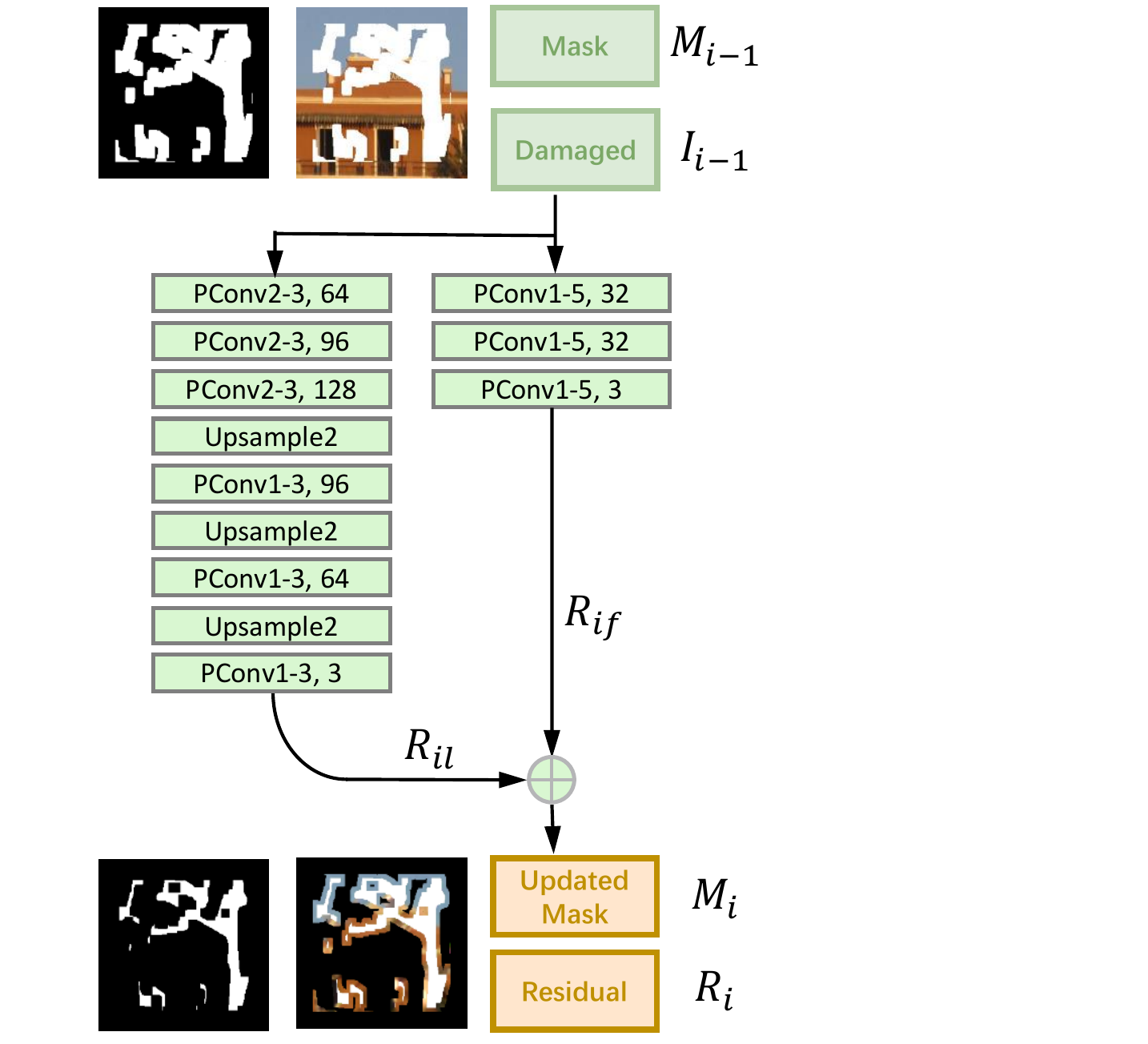}
\caption{Architecture of FRRB, which maintains a full-resolution representation (the right branch) with a conventional feature integration branch (the left branch).} \label{FRRB_arch}
\end{figure}

\subsection{Full-Resolution Residual Block}
Proposed full-resolution residual block (FRRB) focuses on generating residuals in limited area during one dilation step. It is the basic component in final FRRN. Here we first introduce its functionality and then discuss its inference process with full-resolution branch. 

\subsubsection{Functionality of FRRB}
Let us denote $I_{gt}$ as ground truth image, $M$ as original binary mask (0 for the missing areas and 1 for others). Damaged image $I_0$ is synthesized by ground truth image $I_{gt}$ and random-selected mask $M$, \textit{i.e.,} $I_0=I_{gt} \odot M,$ where $\odot$ denotes Hadamard product (element-wise product). We call these areas which are uncovered by masks as clean areas. The functionality of FRRB can be shown as:
\begin{equation}
R_i, M_i= F_i(I_{i-1}, M_{i-1}),\label{equ1}
\end{equation}
\begin{equation}
I_i = I_{i-1} + R_i \odot (1 - M_0),\label{equ2}
\end{equation}
where $F_i$ denotes the i-th FRRB and $R_i$ denotes the residual generated by FRRB. For initializing, $M_0 = M$. Mask can be updated from $M_{i-1}$ to $M_i$ when passing through this FRRB. From equation \ref{equ2}, it is obvious that proposed FRRB generates residual $R_i$ and adds it to its input $I_{i-1}$ after being cropped by original mask , \textit{i.e.},  it will not change values of clean areas.

In short, FRRB provides a residual architecture for progressive image completion. It supports both residual generation and residual structure, which complement each other naturally. Training a deep network which is made of residual blocks has been proved to avoid vanishing gradient problem as well \cite{he2016deep}.

\subsubsection{Specific Architecture of FRRB}

Maintaining high-resolution representations through whole inference process has been proved to be superior for feature fusion \cite{sun2019deep}. The inner architecture of FRRB is also motivated by it and shown in Figure \ref{FRRB_arch}.

As Figure \ref{FRRB_arch} shown, two parallel branches will separately generate residuals. In this figure, PConv $i-j,k$ denotes partial convolution \cite{liu2018image} with stride=i, kernel size=j and channel=k. We hide InstanceNorm \cite{ulyanov2016instance} and activation layers in this figure, which follow every PConv layer. Partial convolution \cite{liu2018image} can be inferred with masks and it is renormalized to be conditioned on only valid pixels. Furthermore, PConv includes a mechanism to automatically generate an updated mask for the next layer as part of the forward pass. The update process of binary masks is similar to classical morphological dilation, which is accomplished naturally by convolution kernels. Here, the shape (hyperparameter) of convolution kernel rather than specific parameter value determines the mask-update process. Therefore, the update process will not be changed during the whole training. Note for binary masks (zero denotes missing area), we implement downsampling in analogous to max-pooling and adopt nearest upsampling. 

In the high-resolution branch, every convolution layer with kernel size equal to 5 will dilate clean areas to extra 2 pixels inside the holes. Three high-resolution layers totally help to dilate clean areas for $2 \times 3=6$ pixels always. In the other branch, input images will be downsampled and upsampled three times for feature extraction and fusion. Every convolution kernel with kernel size 3 will dilate to one more pixel depth into the holes. Three downsampling convolutions and three convolutions with additive upsampling layers can finally dilate for $1 \times 6=6$ pixels as well (sometimes more than 6 pixels because of the asymmetrical process of mask upsampling and downsampling). In short, clean areas will be dilated for 6-pixel width and masks will be eroded for 6 pixels when combining those two branches. Masks updated are denoted as $M_{if}$ in high resolution branch and $M_{il}$ in another branch respectively. Consequently, two branches can both generate residuals and update masks, which can be depicted as:
\begin{equation}
M_i = M_{if} \odot M_{il},\label{equ3}
\end{equation}
\begin{equation}
R_i = \frac{1}{2}(R_{if} \odot M_i + R_{il} \odot M_i).\label{equ4}
\end{equation}
$R_{if}$ and $R_{il}$ denote residuals generated in two branches, which are also described in Figure \ref{FRRB_arch}. Empirically, low resolution branch $M_{il}$ often leads the future integration process for feature prediction and full-resolution branch $M_{if}$ helps to preserve those important features during inference because full-resolution branch will not transform features into different resolutions.

\begin{figure}[t]
    \begin{subfigure}{0.22\columnwidth}
     \includegraphics[width=\textwidth]{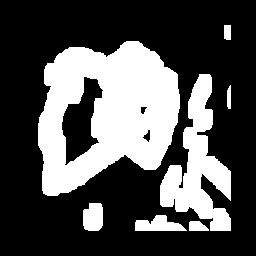}
    \end{subfigure}
    \hspace{0.0015\columnwidth}
    \begin{subfigure}{0.22\columnwidth}
     \includegraphics[width=\textwidth]{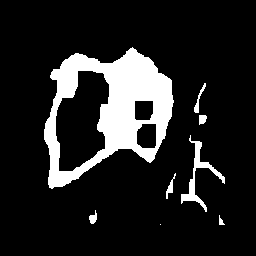}
    \end{subfigure}
    \hspace{0.0015\columnwidth}
    \begin{subfigure}{0.22\columnwidth}
     \includegraphics[width=\textwidth]{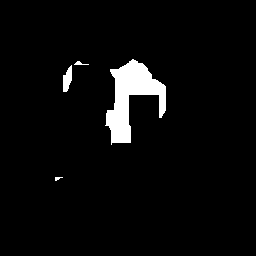}
    \end{subfigure}
    \hspace{0.0015\columnwidth}
    \begin{subfigure}{0.22\columnwidth}
     \includegraphics[width=\textwidth]{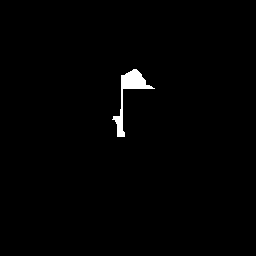}
    \end{subfigure}
    
    \begin{subfigure}{0.22\columnwidth}
     \includegraphics[width=\textwidth]{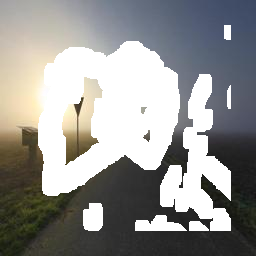}
    \end{subfigure}
    \hspace{0.0015\columnwidth}
    \begin{subfigure}{0.22\columnwidth}
     \includegraphics[width=\textwidth]{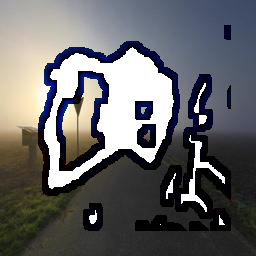}
    \end{subfigure}
    \hspace{0.0015\columnwidth}
    \begin{subfigure}{0.22\columnwidth}
     \includegraphics[width=\textwidth]{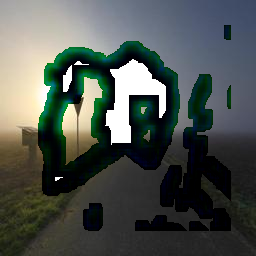}
    \end{subfigure}
    \hspace{0.0015\columnwidth}
    \begin{subfigure}{0.22\columnwidth}
     \includegraphics[width=\textwidth]{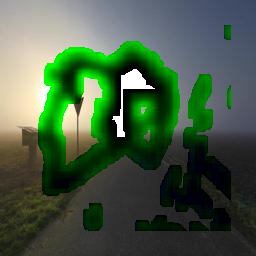}
    \end{subfigure}   
    \caption{Free restoration when stacking FRRBs without adopting \textit{N blocks, one dilation} strategy and step loss function. From left to right: damaged image, output of first FRRB, output of second FRRB, output of third FRRB. It is important to enhance the capability of single dilation module and preserve intermediate reconstruction quality. \label{figure_step_bad}}
\end{figure}

Equation \ref{equ3} and \ref{equ4} explain the function $F$ in Equation \ref{equ1}, which targets at residual generating and mask updating. Noteworthy, in binary matrix $M_{if}$ or binary matrix $M_{il}$, 0 denotes missing region and 1 denotes natural region. As a result, $M_{if} \odot M_{il}$ will generate smaller masks if $M_{if}$ and $M_{il}$ are not the same. 

Input damaged images will be added with generated residuals, which only act on dilated areas (product of updated mask) and will not influence more inner missing pixels. 

Additionally, as missing areas shrink, action scope of FRRB increases progressively. If all missing areas are covered during inpainting, FRRB will degrade to vanilla residual block for image merging. During this period, all pixels inside original missing areas would be refined. 

\subsection{Full-Resolution Residual Network}

Based on designed FRRB, we develop full-resolution residual network (FRRN) for image inpainting with irregular holes. Instead of simply stacking FRRBs to build final network, we adopt \textit{N Blocks, One Dilation} strategy to guarantee the middle restoration quality in one dilation step. 

\subsubsection{\textit{N Blocks, One Dilation} Strategy}

FRRB can accomplish dilation process and erode original holes progressively. However, one FRRB may not be enough for restoring well. Figure \ref{figure_step_bad} shows the middle output when network is simply built by 8 FRRBs. Assuming masks are large enough, this network can help clean areas dilate for eight times. We show several images which respectively are initial damaged image, output after first block, output after second block, output after third block. One FRRB in one dilation step without any limitations will make preceding completion results unacceptable. This may lead to error propagation problem and decrease final performance. 

Consequently, to enhance the ability of each dilation module, it is important to assign more full-resolution residual blocks (FRRBs) in one dilation step. According to experimental results, we finally fix the parameter N to be 2, which will be explained in Section 4.1. Under the setting of $N=2$, completion process in one dilation module can be shown as:
\begin{equation}
R_{i1}, * = F_{i1}(I_{i-1}, M_{i-1}),\label{equ5}
\end{equation}
\begin{equation}
I_{i1} = I_{i-1} + R_{i1} \odot (1 - M_0),
\end{equation}
\begin{equation}
R_{i2}, M_i = F_{i2}(I_{i1}, M_{i-1}),
\end{equation}
\begin{equation}
I_{i2} = I_{i1} + R_{i2} \odot (1 - M_0).
\end{equation}
Then final output of one dilation step is $I_{i} = I_{i2}$, which is calculated by
\begin{equation}
I_{i} = I_{i-1} + (R_{i1} + R_{i2}) \odot (1 - M_0).
\end{equation}
$R_{i1}$ and $R_{i2}$ are generated by two FRRB respectively inside one dilation module. $F_{i1}$ and $F_{i2}$ denote these two residual blocks. Mask will not be updated after the first FRRB, which is represented by * in Equation \ref{equ5}.

In short, by adopting \textit{N Blocks, One Dilation} strategy, where finally we set $N=2$, one dilation module will contain two FRRBs. Each FRRB helps to generate residual in limited region and contributes to high quality of intermediate completion. Masks are unchanged when passing through the first FRRB and will be updated after the second FRRB.

In addition to this \textit{N (Two) Blocks, One Dilation} strategy, we apply a corresponding step loss function to ensure the plausible quality of completion result especially generated by preceding dilation modules, which will also be discussed in Section 3.3.  

\subsubsection{Length of Network}

Utilizing \textit{N (Two) Blocks, One Dilation} strategy, every dilation module with two FRRBs can dilate clean areas for 12 pixels (6 pixels in both sides of holes). If masks are small, one dilation module can cover all missing areas. However, if masks are large, more dilation modules are required. In our experiments, considering both GPU memory requirement and mask size, we fix the dilation step to be 8, \textit{i.e.}, under the setting of N to be 2, the whole number of FRRBs stacked in FRRN is $2 \times 8=16$. We also conduct experiments to show the influence of dilation module's number, which will be illustrated later. 

\subsection{Loss Functions}

Our loss functions measure both pixel-level recovering and global-style similarity. In addition to reconstruction loss, style loss and adversarial loss, we further adopt middle step loss function to guarantee the completion quality after each dilation step. 
 
\subsubsection{Step Loss}

Step loss function contributes to constraining intermediate recovering quality, which is different from four-step reconstruction loss in \cite{zhang2018semantic}. In \cite{zhang2018semantic}, action region of four-step loss function is always stationary due to established dilation method. Step loss for filling irregular holes can adaptively measure the dissimilarity in specified region according to the status of masks. When $I_i$ denotes recovered image output by i-th FRRB and $M_i$ is updated mask, step loss is defined as:
\begin{equation}
L_{step} = \sum_{i=1}^8 \textbf{E} \ \big [ \  \Vert \  (I_i - I_{gt}) \odot M_i \Vert_1 \big ].
\end{equation}
Step loss is accumulated from one to eight because our final FRRN consists of eight dilation modules. Noteworthy, if all missing areas have been covered, step loss degrades to reconstruction loss which acts on the whole region determined by original masks. 

\subsubsection{Reconstruction Loss}

We apply reconstruction loss to constrain pixel-level restoration. $I_{gt}$ is ground truth image, $M$ is given mask. We measure the dissimilarity between recovered image $I_{recover}$ and natural image $I_{gt}$ by adopting $L_1$ norm, which is commonly used and defined as:
\begin{equation}
L_{rec} = \textbf{E} \ \big [ \  \Vert \  I_{gt} - I_{rec} \Vert_1 \big ].
\end{equation}
\subsubsection{Adversarial Loss}

Generative adversarial network (GAN) \cite{goodfellow2014generative} can train a discriminator to judge subjective similarity, which is more closed to human perception. Here, we also utilize adversarial loss to preserve more natural textures and structures in original missing areas, \textit{i.e.}, proposed FRRN is as generator and we train an extra discriminator to minimize adversarial loss, which can be shown as:
\begin{equation}
L_{adv} = \textbf{E} \ \big [ log\ (D\ (I_{gt})) \big ]+ \textbf{E} \ \big [ log\ (1-D\ (I_{rec})) \big ]. 
\end{equation}

In our experiments, we adopt spectral normalization \cite{miyato2018spectral} to stabilize the convergence process of discriminator which is similar with \cite{nazeri2019edgeconnect}. PatchGAN \cite{isola2017image} is also implemented for high-resolution adversarial training. 

\subsubsection{Style Loss}

We also apply style loss \cite{johnson2016perceptual} to ensure that recovered areas are harmonious with background. We perform autocorrelation (Gram matrix) on feature maps computed by VGG19 \cite{simonyan2014very} before applying $L_1$ norm. Features inferred after layer ReLU2-2, ReLU3-4, ReLU4-4, ReLU5-2 will be sent to calculate gram matrix. We denote features of ground truth image and recovering image as $f_{gt}$ and $f_{rec}$. Style loss can be shown as:
\begin{equation}
L_{style} = \textbf{E} \ \big [ \  \Vert G\ (f_{gt}) - G\ (f_{rec}) \Vert_1 \big ],
\end{equation}
where $G$ represents Gram Matrix for computing covariance. In addition to constraining the consistency of whole recovered images, style loss is also proved to be able to eliminate checkerboard artifacts \cite{nazeri2019edgeconnect}. 

In our experiments, we empirically set the weights of these loss functions as:
\begin{equation}
L=20L_{rec}+0.1L_{adv}+100L_{style}+2L_{step}.
\end{equation}

\section{Experiments}

In this section, we start with discussion of experiments which explore the \textit{N Blocks, One Dilation} strategy. Then we compare proposed FRRN with previous promising algorithms \cite{iizuka2017globally,yu2018generative,liu2018image,nazeri2019edgeconnect,DBLP:journals/corr/abs-1903-04227} and show some visual results. Finally, we demonstrate the effectiveness of some critical structures and tricks.

\subsection{Explore \textit{N Blocks, One Dilation} Strategy} Before conducting large-scale experiments of FRRN, we should figure out the suitable value of N in \textit{N Blocks, One Dilation} strategy in advance. 

Figure \ref{figure_step_bad} has shown the unbearable results when network is simply stacked by eight FRRBs. This free restoration would lead to error propagation problem and influence the reconstruction results. We employ \textit{N Blocks, One Dilation} strategy and assign several (N) residual blocks in one dilation module to enhance its capability. Here we conduct two groups of experiments to explore the suitable N value.

\begin{figure}[t]
\centering
\begin{minipage}{\columnwidth}
\begin{subfigure}{\columnwidth}
\includegraphics[scale=0.56, trim={0cm 0 0 0},clip]{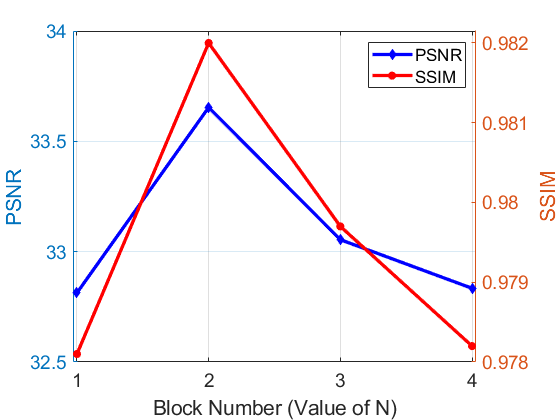}
\caption*{(a) On 10\% to 20\% masks.}
\end{subfigure}
\end{minipage}
\\[11pt]
\begin{minipage}{\columnwidth}
\centering
\resizebox{\columnwidth}{!}{
\begin{tabular}{c|c|c|c|c|c}
    \hline
    & Block Num & 1 & 2 & 3 & 4 \\
    \hline
    \multirow{2}{*}{PSNR} & Mask 10\%-20\% & 32.81 & 33.66 & 33.05 & 32.84 \\
    & Mask 20\%-30\% & 28.31 & 29.01 & 28.83 & 28.78 \\
    \hline
    \multirow{2}{*}{SSIM} & Mask 10\%-20\% & 0.9781 & 0.9820 & 0.9797 & 0.9782 \\
    & Mask 20\%-30\% & 0.9480 & 0.9548 & 0.9538 & 0.9531 \\
    \hline
\end{tabular}}
\caption*{(b) Specific PSNR and SSIM values.}
\caption{Comparison results in terms of PSNR and SSIM for different N values when we evaluate light FRRN on CelebA dataset. N means block number in this figure, which represents how many FRRBs we assign in one dilation module. \label{n_step_effect}}
\end{minipage}
\end{figure}

\begin{figure*}[h]
\centering
\begin{subfigure}{0.16\textwidth}
\includegraphics[scale=0.30]{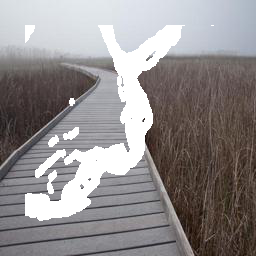}
\end{subfigure}
\begin{subfigure}{0.16\textwidth}
\includegraphics[scale=0.30]{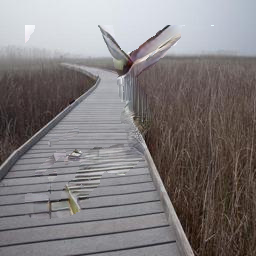}
\end{subfigure}
\begin{subfigure}{0.16\textwidth}
\includegraphics[scale=0.30]{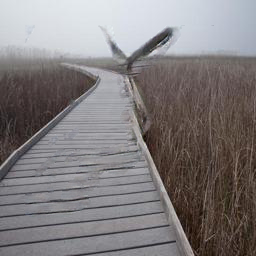}
\end{subfigure}
\begin{subfigure}{0.16\textwidth}
\includegraphics[scale=0.30]{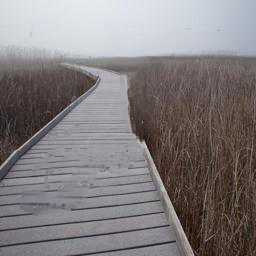}
\end{subfigure}
\begin{subfigure}{0.16\textwidth}
\includegraphics[scale=0.30]{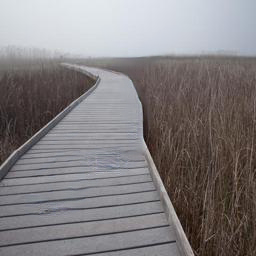}
\end{subfigure}
\begin{subfigure}{0.16\textwidth}
\includegraphics[scale=0.30]{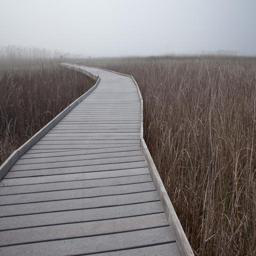}
\end{subfigure}

\begin{subfigure}{0.16\textwidth}
\includegraphics[scale=0.30]{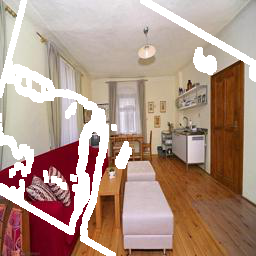}
\caption*{Damaged}
\end{subfigure}
\begin{subfigure}{0.16\textwidth}
\includegraphics[scale=0.30]{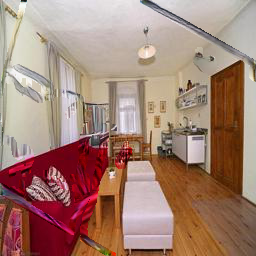}
\caption*{GLCIC\cite{iizuka2017globally}}
\end{subfigure}
\begin{subfigure}{0.16\textwidth}
\includegraphics[scale=0.30]{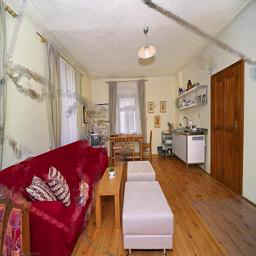}
\caption*{CA\cite{yu2018generative}}
\end{subfigure}
\begin{subfigure}{0.16\textwidth}
\includegraphics[scale=0.30]{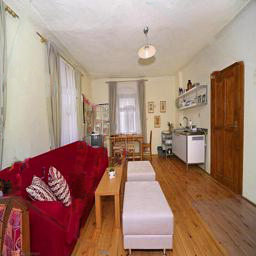}
\caption*{PIC\cite{DBLP:journals/corr/abs-1903-04227}}
\end{subfigure}
\begin{subfigure}{0.16\textwidth}
\includegraphics[scale=0.30]{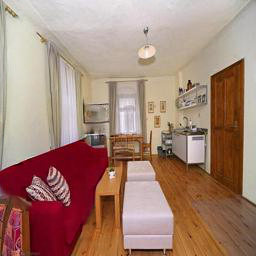}
\caption*{Our FRRN}
\end{subfigure}
\begin{subfigure}{0.16\textwidth}
\includegraphics[scale=0.30]{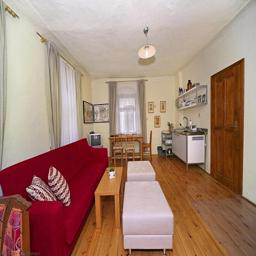}
\caption*{Ground Truth}
\end{subfigure}
\caption{Compare with previous work.\label{figure_compare}}
\end{figure*}

Due to the long training time on Places2 dataset \cite{zhou2017places} and large GPU memory requirement for training the whole FRRN, we select a lighter architecture for this 
exploratory experiment, \textit{i.e.}, we train a light FRRN on CelebA dataset \cite{liu2015deep}, which contains three dilation modules. We select 200000 aligned images for training and the rest 2600 images are for testing.

Considering larger N corresponds to deeper network, for fair comparison, we control number of parameters to be almost the same by decreasing channels of convolution kernels when N is larger. That is to say, we explore the upper bound performance of dilation module that is limited to given parameter number. We choose two categories of masks in this part. 10\%-20\% masks denote those masks whose holes cover 10\% to 20\% pixels when synthesizing damaged images. 20\%-30\% masks mean the holes are larger. In our experiments, we utilize the mask dataset released by \cite{liu2018image}, which contains 12000 irregular masks with holes of various sizes.

We draw a figure and list the specific Peak Signal to Noise Ratio (PSNR) value and Structural Similarity Index (SSIM) value \cite{wang2004image} (window size is 11 in the whole experiments) to explore the influence of N (Figure \ref{n_step_effect}). As Figure \ref{n_step_effect} shown, \textit{Two Blocks, One Dilation} strategy performs the best even when N is larger. Useless stacking FRRB in one dilation module makes no obvious improvement. Hence, we determine $N=2$ in our \textit{N Blocks, One Dilation} strategy, then our final FRRN is supposed to have 16 FRRBs (eight dilation modules assigned and each one contains two FRRBs).

\subsection{Comparison}

We first conduct large-scale experiments to evaluate proposed FRRN on Places2 \cite{zhou2017places}. We train FRRN on its training data and test on its whole 36500 validation images. For fair comparison, we resize images and masks to $256\times256$ and then send them into network, which follows the setting of \cite{nazeri2019edgeconnect}. Note proposed FRRN can deal with different input size as long as input resolution is an integer multiple of eight because every FRRB will down/upsample features for three times. 

\subsubsection{Training Details}
We train our final FRRN which contains 16 FRRBs (eight dilation modules, each contains two FRRBs). We use two Tesla V100 GPUs (16GB) to train our network with batch size equal to 8. As for training one whole epoch, it takes about 42 hours in Places2 \cite{zhou2017places}. We adopt Adam optimizer with $\beta_1 = 0$,  $\beta_2=0.9$, and set learning rate for training generator and discriminator as 10:1. We train one single network to deal with different masks, \textit{i.e}., all 12000 masks released by \cite{liu2018image} with various-sized holes are randomly picked up for training.

\subsubsection{Qualitative Comparison}

\begin{figure}[t]
\centering
\begin{subfigure}{0.24\columnwidth}
\includegraphics[scale=0.215]{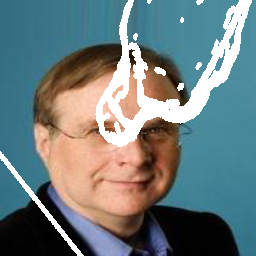}
\end{subfigure}
\begin{subfigure}{0.24\columnwidth}
\includegraphics[scale=0.215]{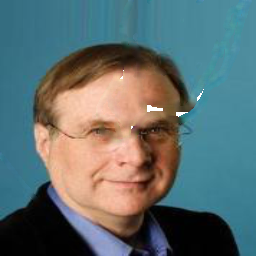}
\end{subfigure}
\begin{subfigure}{0.24\columnwidth}
\includegraphics[scale=0.215]{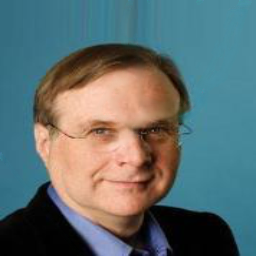}
\end{subfigure}
\begin{subfigure}{0.24\columnwidth}
\includegraphics[scale=0.215]{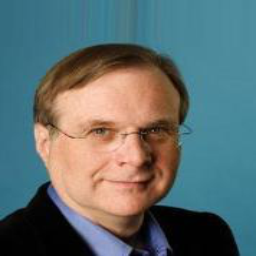}
\end{subfigure}

\begin{subfigure}{0.24\columnwidth}
\includegraphics[scale=0.215]{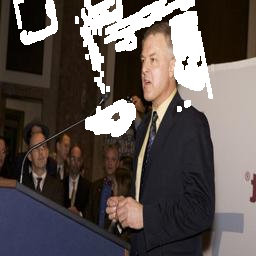}
\end{subfigure}
\begin{subfigure}{0.24\columnwidth}
\includegraphics[scale=0.215]{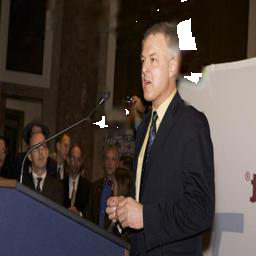}
\end{subfigure}
\begin{subfigure}{0.24\columnwidth}
\includegraphics[scale=0.215]{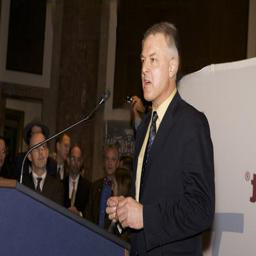}
\end{subfigure}
\begin{subfigure}{0.24\columnwidth}
\includegraphics[scale=0.215]{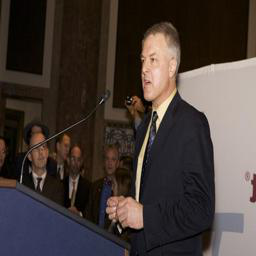}
\end{subfigure}

\begin{subfigure}{0.24\columnwidth}
\includegraphics[scale=0.215]{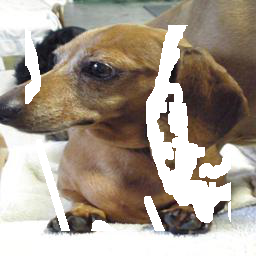}
\end{subfigure}
\begin{subfigure}{0.24\columnwidth}
\includegraphics[scale=0.215]{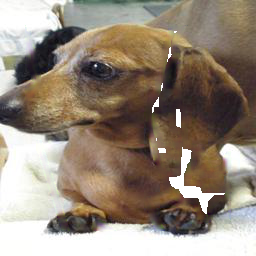}
\end{subfigure}
\begin{subfigure}{0.24\columnwidth}
\includegraphics[scale=0.215]{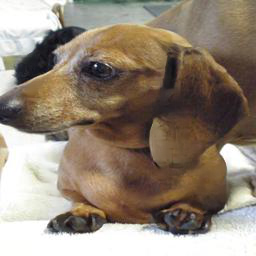}
\end{subfigure}
\begin{subfigure}{0.24\columnwidth}
\includegraphics[scale=0.215]{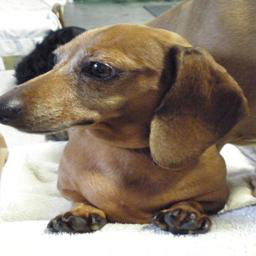}
\end{subfigure}

\begin{subfigure}{0.24\columnwidth}
\includegraphics[scale=0.215]{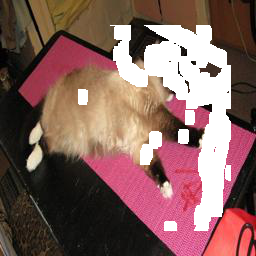}
\end{subfigure}
\begin{subfigure}{0.24\columnwidth}
\includegraphics[scale=0.215]{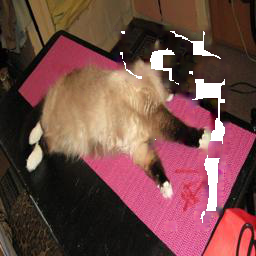}
\end{subfigure}
\begin{subfigure}{0.24\columnwidth}
\includegraphics[scale=0.215]{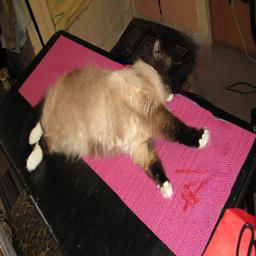}
\end{subfigure}
\begin{subfigure}{0.24\columnwidth}
\includegraphics[scale=0.215]{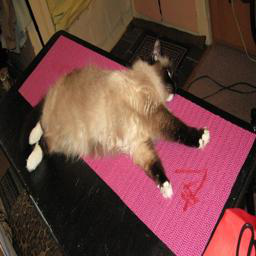}
\end{subfigure}

\begin{subfigure}{0.24\columnwidth}
\includegraphics[scale=0.215]{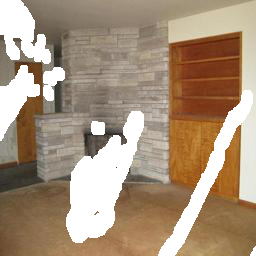}
\end{subfigure}
\begin{subfigure}{0.24\columnwidth}
\includegraphics[scale=0.215]{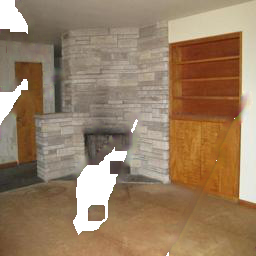}
\end{subfigure}
\begin{subfigure}{0.24\columnwidth}
\includegraphics[scale=0.215]{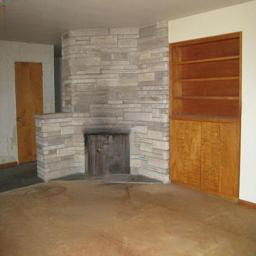}
\end{subfigure}
\begin{subfigure}{0.24\columnwidth}
\includegraphics[scale=0.215]{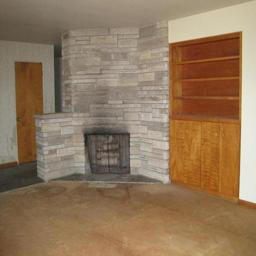}
\end{subfigure}

\begin{subfigure}{0.24\columnwidth}
\includegraphics[scale=0.215]{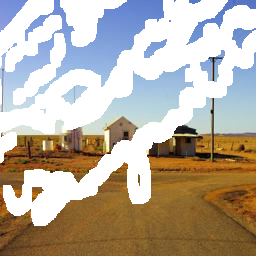}
\end{subfigure}
\begin{subfigure}{0.24\columnwidth}
\includegraphics[scale=0.215]{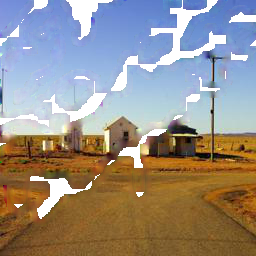}
\end{subfigure}
\begin{subfigure}{0.24\columnwidth}
\includegraphics[scale=0.215]{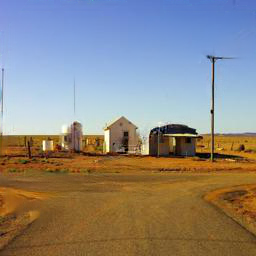}
\end{subfigure}
\begin{subfigure}{0.24\columnwidth}
\includegraphics[scale=0.215]{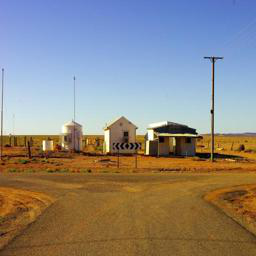}
\end{subfigure}
\caption{From left to right: damaged image, coarse restoration after one dilation module, final restoration, ground truth. \label{More_results}}
\end{figure}

Figure \ref{figure_compare} shows the visual results of FRRN compared with several latest methods such as GLCIC \cite{iizuka2017globally}, CA \cite{yu2018generative}, PIC \cite{DBLP:journals/corr/abs-1903-04227}. These results demonstrate that though we do not design a separated network for edge generation \cite{DBLP:journals/corr/abs-1903-04227}, recovered images still preserve plausible structures. It is reasonable when we relate proposed progressive framework with traditional dilation operator. In classical morphological processing, dilation then erosion will help to connect broken contours. Proposed method utilizing residual architecture can update irregular masks step by step and enable convolution kernels to complete contours.  

More results of FRRN are presented in Figure \ref{More_results}. We put an extra column to show the middle output when damaged image is preliminarily recovered by the first dilation module. With the constraint of \textit{Two Blocks, One Dilation} strategy and step loss function, intermediate reconstructions are also satisfactory. The top two lines in Figure \ref{More_results} are randomly selected from CelebA \cite{liu2015deep} dataset. They prove that proposed FRRN performs well not only on simple-structured images, which occupy the majority on Places2 \cite{zhou2017places}, but also on complicated images such as portraits. Though occasionally FRRN will generate blurriness, the generated blurriness is still compatible with background of clean areas.

\subsubsection{Quantitative Comparison}

When it comes to quantitative comparison, the superiority of proposed FRRN is more obvious. 

\begin{table}[t]
\centering
\resizebox{\columnwidth}{!}{
\begin{tabular}{c|c|c|c|c|c|c}
    \hline
    \hline
    & Masks & GLCIP & CA & PConv* & EC & Ours \\
    \hline
    \multirow{4}{*}{PSNR$^+$}
     & 10\%-20\% & 23.40 & 24.44 & 28.02 & 27.60 & \textbf{29.86} \\
     & 20\%-30\% & 20.45 & 21.34 & 24.90 & 24.65 & \textbf{26.29} \\
     & 30\%-40\% & 18.38 & 19.31 & 22.45 & 22.64 & \textbf{24.01} \\
     & 40\%-50\% & 16.99 & 17.90 & 20.86 & 21.00 & \textbf{22.08} \\
    \hline
    \multirow{4}{*}{SSIM$^+$}
     & 10\%-20\% & 0.855 & 0.886 & 0.869 & 0.919 & \textbf{0.948} \\
     & 20\%-30\% & 0.762 & 0.806 & 0.777 & 0.857 & \textbf{0.898} \\
     & 30\%-40\% & 0.672 & 0.725 & 0.685 & 0.792 & \textbf{0.842} \\
     & 40\%-50\% & 0.585 & 0.644 & 0.589 & 0.720 & \textbf{0.776} \\
    \hline
    \multirow{4}{*}{$L_{1}(\%)^-$}
     & 10\%-20\% & 2.31 & 2.05 & 1.14 & 1.32 & \textbf{0.82} \\
     & 20\%-30\% & 3.94 & 3.50 & 1.98 & 2.25 & \textbf{1.56} \\
     & 30\%-40\% & 5.73 & 5.04 & 3.02 & 3.23 & \textbf{2.38} \\
     & 40\%-50\% & 7.52 & 6.62 & 4.11 & 4.38 & \textbf{3.39} \\
    \hline
    \hline
\end{tabular}}
\caption{Quantitative results (PSNR, SSIM, $L_1$) of different methods such as GLCIP \cite{iizuka2017globally}, CA \cite{yu2018generative}, PConv \cite{liu2018image} and EC \cite{nazeri2019edgeconnect}). - denotes lower is better and + denotes higher is better. * means that statistics are obtained from their paper. }
\label{Quantitativetabel}
\end{table}

Table \ref{Quantitativetabel} presents the specific values. In this table, GLCIP denotes Global and Local Consistent Image Completion \cite{iizuka2017globally}, CA denotes Contextual Attention \cite{yu2018generative}, PConv is Partial Convolution Unet \cite{liu2018image} and EC means EdgeConnect \cite{nazeri2019edgeconnect}, which is previous state-of-the-art method. When measure the reconstruction quality by PSNR, SSIM \cite{wang2004image} (window size is also 11) and $L_1$ distance. Our FRRN performs much better than other methods. 

\subsection{Ablation Study}

Here we conduct experiments to show the effectiveness of full-resolution branch and step loss function. We also explore the influence of dilation module's number in this section.

\subsubsection{Effective Residual Block: FRRB}

Proposed FRRB described in Section 3.1 maintains a full-resolution branch for feature fusion. We conduct an experiment to explore the effectiveness of  full-resolution branch. We replace all FRRBs in light FRRN (containing three dilation modules with six FRRBs) to conventional single-channel residual block, which also consists of several three downsampling and three upsampling without this full-resolution branch. Other parts in light FRRN are unchanged such as training strategy and loss function. Experiments in this part are conducted on CelebA \cite{liu2015deep} with masks covering 10\%-20\% regions.

\begin{figure}[t]
    \begin{minipage}{0.57\linewidth}
    \includegraphics[scale=0.36, trim={0.25cm 0 0 0.3},clip]{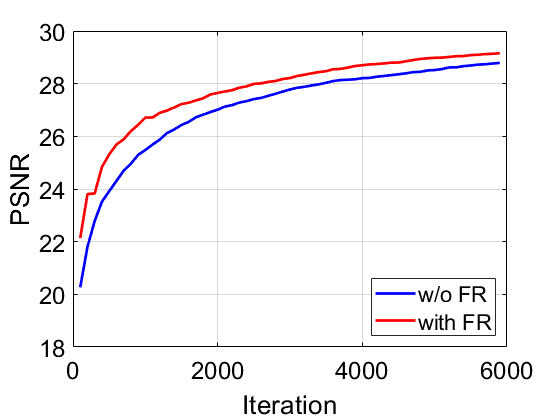}
    \end{minipage}
    \begin{minipage}{0.38\linewidth}
    \centering
    \scalebox{0.86}{
    \begin{tabular}{|c|c|c|}
        \hline
         & PSNR & SSIM \\
        \hline
        with FR & 33.66 & 0.982\\
        \hline
        w/o FR & 33.03 & 0.980\\
        \hline
    \end{tabular}}    
    \end{minipage}
    \caption{Experiments on light FRRN with or without full-resolution branch. Full-resolution (FR) branch benefits both network convergence and final performance. \label{time_compare}}
\end{figure}

Figure \ref{time_compare} shows the convergence speed with or without full-resolution branch during the initial 6000 iterations. It verifies that maintaining a full-resolution branch accelerates convergence speed and can achieve higher PSNR. 

\subsubsection{Performance of Step Loss}

Step loss adopted here is adaptive to deal with different irregular masks. Corresponding to this progressive inpainting approach, we impose step loss to constrain intermediate reconstruction quality. Here we also perform an ablation study to prove that step loss benefits restoration quality. 

We train light FRRN with or without step loss in different situations. Table \ref{table3} shows the difference of PSNR. No matter what the size of mask is, step loss improves the restoration results. Experimental results behave the same as expectations because free restoration without step loss will arouse error propagation problem, which has been presented in Figure \ref{figure_step_bad}. 

\begin{table}[t]
\centering
\resizebox{\columnwidth}{!}{
\begin{tabular}{c|c|c|c}
    \hline
    Mask Size & 10\%-20\% & 20\%-30\% & 30\%-40\% \\
    \hline
    with step loss & 33.66 & 29.01 & 25.05\\
    \hline
    w/o step loss & 33.08 & 28.87 & 24.76\\
    \hline
\end{tabular}}
\caption{PSNR of light FRRN trained with or without step loss. Evaluation is performed on CelebA.}
\label{table3}
\end{table}

\subsubsection{Influence of Network Length}

Proposed FRRB provides a way that enables inpainting network to be deeper. Mentioned light FRRN which contains three dilation modules can fill holes with 36 pixels in diameter (Each dilation module can dilate clean regions for 12 pixels in diameter). 

\begin{figure}[t]
    \begin{minipage}{0.54\linewidth}
     \includegraphics[scale=0.36, trim={0.2cm 0cm 0.5cm 0.5cm},clip]{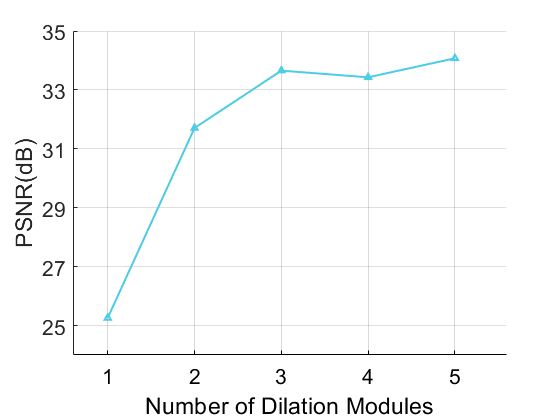}
    \end{minipage}
    \hspace{0.0015\linewidth}
    \begin{minipage}{0.44\linewidth}
    \scalebox{0.85}{
    \begin{tabular}{c|c|c}
        \hline
        Number & PSNR & SSIM \\
        \hline
        1 & 25.36 & 0.9081\\
        \hline
        2 & 31.71 & 0.9714\\
        \hline
        3 & 33.66 & 0.9820\\
        \hline
        4 & 33.43 & 0.9810\\
        \hline
        5 & 34.07 & 0.9831\\
        \hline
    \end{tabular}}
    \end{minipage}
    \caption{Influence of dilation module's number. We conduct experiments on 10\%-20\% masks. Three dilation modules are almost enough to reach the peak because holes are relatively small.}
    \label{dilation_num}
\end{figure}

We further explore the influence of dilation module number. In Figure \ref{dilation_num}, we could observe that when complete images with mask size from 10\% to 20\%, network with three dilation modules can almost reach the peak performance. 

However, when masks are large, three dilation steps are not enough to handle damaged images. More dilation modules are required. We know two FRRBs in one module can achieve better performance. Our final FRRN consists of eight modules which is determined empirically, \textit{i.e.}, there are 16 FRRBs in our network. Such network can deal with holes up to 96 pixels in diameter and is exactly suitable to train on two Tesla V100 GPUs. 

\section{Conclusion}

In this paper, we propose a novel full-resolution residual network (FRRN), which helps us achieve the state-of-the-art image inpainting results. Full-resolution residual block (FRRB), a component of FRRN, maintains a full-resolution representation which facilitates both texture reconstruction and network convergence. This residual framework is particularly suitable for progressive image completion. Besides, in order to \textit{restore well and then update masks}, we adopt \textit{N (Two) Blocks, One Dilation} strategy and apply step loss function to guarantee the intermediate reconstruction quality. Proposed FRRN provides an effective method for progressively filling irregular holes. However, when holes are huge, the required number of parameters will raise up obviously. Increasing the dilation stride or decreasing parameter number in one dilation module would be feasible, which we regard as future work.

\begin{acks}
This work was supported in part by NSFC under Grant 61571413, 61632001.
\end{acks}

%
\bibliographystyle{ACM-Reference-Format}
\bibliography{sig_conf}


\begin{thebibliography}{36}


\ifx \showCODEN    \undefined \def \showCODEN     #1{\unskip}     \fi
\ifx \showDOI      \undefined \def \showDOI       #1{#1}\fi
\ifx \showISBNx    \undefined \def \showISBNx     #1{\unskip}     \fi
\ifx \showISBNxiii \undefined \def \showISBNxiii  #1{\unskip}     \fi
\ifx \showISSN     \undefined \def \showISSN      #1{\unskip}     \fi
\ifx \showLCCN     \undefined \def \showLCCN      #1{\unskip}     \fi
\ifx \shownote     \undefined \def \shownote      #1{#1}          \fi
\ifx \showarticletitle \undefined \def \showarticletitle #1{#1}   \fi
\ifx \showURL      \undefined \def \showURL       {\relax}        \fi
\providecommand\bibfield[2]{#2}
\providecommand\bibinfo[2]{#2}
\providecommand\natexlab[1]{#1}
\providecommand\showeprint[2][]{arXiv:#2}

\bibitem[\protect\citeauthoryear{Ballester, Bertalmio, Caselles, Sapiro, and
  Verdera}{Ballester et~al\mbox{.}}{2000}]%
        {ballester2000filling}
\bibfield{author}{\bibinfo{person}{Coloma Ballester}, \bibinfo{person}{Marcelo
  Bertalmio}, \bibinfo{person}{Vicent Caselles}, \bibinfo{person}{Guillermo
  Sapiro}, {and} \bibinfo{person}{Joan Verdera}.}
  \bibinfo{year}{2000}\natexlab{}.
\newblock \showarticletitle{Filling-in by joint interpolation of vector fields
  and gray levels}.
\newblock  (\bibinfo{year}{2000}).
\newblock


\bibitem[\protect\citeauthoryear{Barnes, Shechtman, Finkelstein, and
  Goldman}{Barnes et~al\mbox{.}}{2009}]%
        {barnes2009patchmatch}
\bibfield{author}{\bibinfo{person}{Connelly Barnes}, \bibinfo{person}{Eli
  Shechtman}, \bibinfo{person}{Adam Finkelstein}, {and} \bibinfo{person}{Dan~B
  Goldman}.} \bibinfo{year}{2009}\natexlab{}.
\newblock \showarticletitle{PatchMatch: A randomized correspondence algorithm
  for structural image editing}. In \bibinfo{booktitle}{\emph{ACM Transactions
  on Graphics (ToG)}}, Vol.~\bibinfo{volume}{28}. ACM, \bibinfo{pages}{24}.
\newblock


\bibitem[\protect\citeauthoryear{Bertalmio, Sapiro, Caselles, and
  Ballester}{Bertalmio et~al\mbox{.}}{2000}]%
        {bertalmio2000image}
\bibfield{author}{\bibinfo{person}{Marcelo Bertalmio},
  \bibinfo{person}{Guillermo Sapiro}, \bibinfo{person}{Vincent Caselles}, {and}
  \bibinfo{person}{Coloma Ballester}.} \bibinfo{year}{2000}\natexlab{}.
\newblock \showarticletitle{Image inpainting}. In
  \bibinfo{booktitle}{\emph{Proceedings of the 27th annual conference on
  Computer graphics and interactive techniques}}. ACM Press/Addison-Wesley
  Publishing Co., \bibinfo{pages}{417--424}.
\newblock


\bibitem[\protect\citeauthoryear{Bertalmio, Vese, Sapiro, and Osher}{Bertalmio
  et~al\mbox{.}}{2003}]%
        {bertalmio2003simultaneous}
\bibfield{author}{\bibinfo{person}{Marcelo Bertalmio},
  \bibinfo{person}{Luminita Vese}, \bibinfo{person}{Guillermo Sapiro}, {and}
  \bibinfo{person}{Stanley Osher}.} \bibinfo{year}{2003}\natexlab{}.
\newblock \showarticletitle{Simultaneous structure and texture image
  inpainting}.
\newblock \bibinfo{journal}{\emph{IEEE transactions on image processing}}
  \bibinfo{volume}{12}, \bibinfo{number}{8} (\bibinfo{year}{2003}),
  \bibinfo{pages}{882--889}.
\newblock


\bibitem[\protect\citeauthoryear{Darabi, Shechtman, Barnes, Goldman, and
  Sen}{Darabi et~al\mbox{.}}{2012}]%
        {darabi2012image}
\bibfield{author}{\bibinfo{person}{Soheil Darabi}, \bibinfo{person}{Eli
  Shechtman}, \bibinfo{person}{Connelly Barnes}, \bibinfo{person}{Dan~B
  Goldman}, {and} \bibinfo{person}{Pradeep Sen}.}
  \bibinfo{year}{2012}\natexlab{}.
\newblock \showarticletitle{Image melding: Combining inconsistent images using
  patch-based synthesis.}
\newblock \bibinfo{journal}{\emph{ACM Trans. Graph.}} \bibinfo{volume}{31},
  \bibinfo{number}{4} (\bibinfo{year}{2012}), \bibinfo{pages}{82--1}.
\newblock


\bibitem[\protect\citeauthoryear{Drori, Cohen-Or, and Yeshurun}{Drori
  et~al\mbox{.}}{2003}]%
        {drori2003fragment}
\bibfield{author}{\bibinfo{person}{Iddo Drori}, \bibinfo{person}{Daniel
  Cohen-Or}, {and} \bibinfo{person}{Hezy Yeshurun}.}
  \bibinfo{year}{2003}\natexlab{}.
\newblock \showarticletitle{Fragment-based image completion}. In
  \bibinfo{booktitle}{\emph{ACM Transactions on graphics (TOG)}},
  Vol.~\bibinfo{volume}{22}. ACM, \bibinfo{pages}{303--312}.
\newblock


\bibitem[\protect\citeauthoryear{Esedoglu and Shen}{Esedoglu and Shen}{2002}]%
        {esedoglu2002digital}
\bibfield{author}{\bibinfo{person}{Selim Esedoglu} {and}
  \bibinfo{person}{Jianhong Shen}.} \bibinfo{year}{2002}\natexlab{}.
\newblock \showarticletitle{Digital inpainting based on the
  Mumford--Shah--Euler image model}.
\newblock \bibinfo{journal}{\emph{European Journal of Applied Mathematics}}
  \bibinfo{volume}{13}, \bibinfo{number}{4} (\bibinfo{year}{2002}),
  \bibinfo{pages}{353--370}.
\newblock


\bibitem[\protect\citeauthoryear{Goodfellow, Pouget-Abadie, Mirza, Xu,
  Warde-Farley, Ozair, Courville, and Bengio}{Goodfellow et~al\mbox{.}}{2014}]%
        {goodfellow2014generative}
\bibfield{author}{\bibinfo{person}{Ian Goodfellow}, \bibinfo{person}{Jean
  Pouget-Abadie}, \bibinfo{person}{Mehdi Mirza}, \bibinfo{person}{Bing Xu},
  \bibinfo{person}{David Warde-Farley}, \bibinfo{person}{Sherjil Ozair},
  \bibinfo{person}{Aaron Courville}, {and} \bibinfo{person}{Yoshua Bengio}.}
  \bibinfo{year}{2014}\natexlab{}.
\newblock \showarticletitle{Generative adversarial nets}. In
  \bibinfo{booktitle}{\emph{Advances in neural information processing
  systems}}. \bibinfo{pages}{2672--2680}.
\newblock


\bibitem[\protect\citeauthoryear{He, Zhang, Ren, and Sun}{He
  et~al\mbox{.}}{2016}]%
        {he2016deep}
\bibfield{author}{\bibinfo{person}{Kaiming He}, \bibinfo{person}{Xiangyu
  Zhang}, \bibinfo{person}{Shaoqing Ren}, {and} \bibinfo{person}{Jian Sun}.}
  \bibinfo{year}{2016}\natexlab{}.
\newblock \showarticletitle{Deep residual learning for image recognition}. In
  \bibinfo{booktitle}{\emph{Proceedings of the IEEE conference on computer
  vision and pattern recognition}}. \bibinfo{pages}{770--778}.
\newblock


\bibitem[\protect\citeauthoryear{Huang, Kang, Ahuja, and Kopf}{Huang
  et~al\mbox{.}}{2014}]%
        {huang2014image}
\bibfield{author}{\bibinfo{person}{Jia-Bin Huang}, \bibinfo{person}{Sing~Bing
  Kang}, \bibinfo{person}{Narendra Ahuja}, {and} \bibinfo{person}{Johannes
  Kopf}.} \bibinfo{year}{2014}\natexlab{}.
\newblock \showarticletitle{Image completion using planar structure guidance}.
\newblock \bibinfo{journal}{\emph{ACM Transactions on graphics (TOG)}}
  \bibinfo{volume}{33}, \bibinfo{number}{4} (\bibinfo{year}{2014}),
  \bibinfo{pages}{129}.
\newblock


\bibitem[\protect\citeauthoryear{Iizuka, Simo-Serra, and Ishikawa}{Iizuka
  et~al\mbox{.}}{2017}]%
        {iizuka2017globally}
\bibfield{author}{\bibinfo{person}{Satoshi Iizuka}, \bibinfo{person}{Edgar
  Simo-Serra}, {and} \bibinfo{person}{Hiroshi Ishikawa}.}
  \bibinfo{year}{2017}\natexlab{}.
\newblock \showarticletitle{Globally and locally consistent image completion}.
\newblock \bibinfo{journal}{\emph{ACM Transactions on Graphics (ToG)}}
  \bibinfo{volume}{36}, \bibinfo{number}{4} (\bibinfo{year}{2017}),
  \bibinfo{pages}{107}.
\newblock


\bibitem[\protect\citeauthoryear{Isola, Zhu, Zhou, and Efros}{Isola
  et~al\mbox{.}}{2017}]%
        {isola2017image}
\bibfield{author}{\bibinfo{person}{Phillip Isola}, \bibinfo{person}{Jun-Yan
  Zhu}, \bibinfo{person}{Tinghui Zhou}, {and} \bibinfo{person}{Alexei~A
  Efros}.} \bibinfo{year}{2017}\natexlab{}.
\newblock \showarticletitle{Image-to-image translation with conditional
  adversarial networks}. In \bibinfo{booktitle}{\emph{Proceedings of the IEEE
  conference on computer vision and pattern recognition}}.
  \bibinfo{pages}{1125--1134}.
\newblock


\bibitem[\protect\citeauthoryear{Johnson, Alahi, and Fei-Fei}{Johnson
  et~al\mbox{.}}{2016}]%
        {johnson2016perceptual}
\bibfield{author}{\bibinfo{person}{Justin Johnson}, \bibinfo{person}{Alexandre
  Alahi}, {and} \bibinfo{person}{Li Fei-Fei}.} \bibinfo{year}{2016}\natexlab{}.
\newblock \showarticletitle{Perceptual losses for real-time style transfer and
  super-resolution}. In \bibinfo{booktitle}{\emph{European conference on
  computer vision}}. Springer, \bibinfo{pages}{694--711}.
\newblock


\bibitem[\protect\citeauthoryear{Kim, Kwon~Lee, and Mu~Lee}{Kim
  et~al\mbox{.}}{2016}]%
        {kim2016accurate}
\bibfield{author}{\bibinfo{person}{Jiwon Kim}, \bibinfo{person}{Jung Kwon~Lee},
  {and} \bibinfo{person}{Kyoung Mu~Lee}.} \bibinfo{year}{2016}\natexlab{}.
\newblock \showarticletitle{Accurate image super-resolution using very deep
  convolutional networks}. In \bibinfo{booktitle}{\emph{Proceedings of the IEEE
  conference on computer vision and pattern recognition}}.
  \bibinfo{pages}{1646--1654}.
\newblock


\bibitem[\protect\citeauthoryear{Levin, Zomet, and Weiss}{Levin
  et~al\mbox{.}}{2003}]%
        {levin2003learning}
\bibfield{author}{\bibinfo{person}{Anat Levin}, \bibinfo{person}{Assaf Zomet},
  {and} \bibinfo{person}{Yair Weiss}.} \bibinfo{year}{2003}\natexlab{}.
\newblock \showarticletitle{Learning how to inpaint from global image
  statistics}. In \bibinfo{booktitle}{\emph{null}}. IEEE, \bibinfo{pages}{305}.
\newblock


\bibitem[\protect\citeauthoryear{Lim, Son, Kim, Nah, and Mu~Lee}{Lim
  et~al\mbox{.}}{2017}]%
        {lim2017enhanced}
\bibfield{author}{\bibinfo{person}{Bee Lim}, \bibinfo{person}{Sanghyun Son},
  \bibinfo{person}{Heewon Kim}, \bibinfo{person}{Seungjun Nah}, {and}
  \bibinfo{person}{Kyoung Mu~Lee}.} \bibinfo{year}{2017}\natexlab{}.
\newblock \showarticletitle{Enhanced deep residual networks for single image
  super-resolution}. In \bibinfo{booktitle}{\emph{Proceedings of the IEEE
  Conference on Computer Vision and Pattern Recognition Workshops}}.
  \bibinfo{pages}{136--144}.
\newblock


\bibitem[\protect\citeauthoryear{Liu, Sun, Wu, Li, and Zhang}{Liu
  et~al\mbox{.}}{2007}]%
        {liu2007image}
\bibfield{author}{\bibinfo{person}{Dong Liu}, \bibinfo{person}{Xiaoyan Sun},
  \bibinfo{person}{Feng Wu}, \bibinfo{person}{Shipeng Li}, {and}
  \bibinfo{person}{Ya-Qin Zhang}.} \bibinfo{year}{2007}\natexlab{}.
\newblock \showarticletitle{Image compression with edge-based inpainting}.
\newblock \bibinfo{journal}{\emph{IEEE Transactions on Circuits and Systems for
  Video Technology}} \bibinfo{volume}{17}, \bibinfo{number}{10}
  (\bibinfo{year}{2007}), \bibinfo{pages}{1273--1287}.
\newblock


\bibitem[\protect\citeauthoryear{Liu, Reda, Shih, Wang, Tao, and Catanzaro}{Liu
  et~al\mbox{.}}{2018}]%
        {liu2018image}
\bibfield{author}{\bibinfo{person}{Guilin Liu}, \bibinfo{person}{Fitsum~A
  Reda}, \bibinfo{person}{Kevin~J Shih}, \bibinfo{person}{Ting-Chun Wang},
  \bibinfo{person}{Andrew Tao}, {and} \bibinfo{person}{Bryan Catanzaro}.}
  \bibinfo{year}{2018}\natexlab{}.
\newblock \showarticletitle{Image inpainting for irregular holes using partial
  convolutions}. In \bibinfo{booktitle}{\emph{Proceedings of the European
  Conference on Computer Vision (ECCV)}}. \bibinfo{pages}{85--100}.
\newblock


\bibitem[\protect\citeauthoryear{Liu, Luo, Wang, and Tang}{Liu
  et~al\mbox{.}}{2015}]%
        {liu2015deep}
\bibfield{author}{\bibinfo{person}{Ziwei Liu}, \bibinfo{person}{Ping Luo},
  \bibinfo{person}{Xiaogang Wang}, {and} \bibinfo{person}{Xiaoou Tang}.}
  \bibinfo{year}{2015}\natexlab{}.
\newblock \showarticletitle{Deep learning face attributes in the wild}. In
  \bibinfo{booktitle}{\emph{Proceedings of the IEEE international conference on
  computer vision}}. \bibinfo{pages}{3730--3738}.
\newblock


\bibitem[\protect\citeauthoryear{Miyato, Kataoka, Koyama, and Yoshida}{Miyato
  et~al\mbox{.}}{2018}]%
        {miyato2018spectral}
\bibfield{author}{\bibinfo{person}{Takeru Miyato}, \bibinfo{person}{Toshiki
  Kataoka}, \bibinfo{person}{Masanori Koyama}, {and} \bibinfo{person}{Yuichi
  Yoshida}.} \bibinfo{year}{2018}\natexlab{}.
\newblock \showarticletitle{Spectral normalization for generative adversarial
  networks}.
\newblock \bibinfo{journal}{\emph{arXiv preprint arXiv:1802.05957}}
  (\bibinfo{year}{2018}).
\newblock


\bibitem[\protect\citeauthoryear{Nazeri, Ng, Joseph, Qureshi, and
  Ebrahimi}{Nazeri et~al\mbox{.}}{2019}]%
        {nazeri2019edgeconnect}
\bibfield{author}{\bibinfo{person}{Kamyar Nazeri}, \bibinfo{person}{Eric Ng},
  \bibinfo{person}{Tony Joseph}, \bibinfo{person}{Faisal Qureshi}, {and}
  \bibinfo{person}{Mehran Ebrahimi}.} \bibinfo{year}{2019}\natexlab{}.
\newblock \showarticletitle{EdgeConnect: Generative Image Inpainting with
  Adversarial Edge Learning}.
\newblock \bibinfo{journal}{\emph{arXiv preprint arXiv:1901.00212}}
  (\bibinfo{year}{2019}).
\newblock


\bibitem[\protect\citeauthoryear{Pathak, Krahenbuhl, Donahue, Darrell, and
  Efros}{Pathak et~al\mbox{.}}{2016}]%
        {pathak2016context}
\bibfield{author}{\bibinfo{person}{Deepak Pathak}, \bibinfo{person}{Philipp
  Krahenbuhl}, \bibinfo{person}{Jeff Donahue}, \bibinfo{person}{Trevor
  Darrell}, {and} \bibinfo{person}{Alexei~A Efros}.}
  \bibinfo{year}{2016}\natexlab{}.
\newblock \showarticletitle{Context encoders: Feature learning by inpainting}.
  In \bibinfo{booktitle}{\emph{Proceedings of the IEEE conference on computer
  vision and pattern recognition}}. \bibinfo{pages}{2536--2544}.
\newblock


\bibitem[\protect\citeauthoryear{Simonyan and Zisserman}{Simonyan and
  Zisserman}{2014}]%
        {simonyan2014very}
\bibfield{author}{\bibinfo{person}{Karen Simonyan} {and}
  \bibinfo{person}{Andrew Zisserman}.} \bibinfo{year}{2014}\natexlab{}.
\newblock \showarticletitle{Very deep convolutional networks for large-scale
  image recognition}.
\newblock \bibinfo{journal}{\emph{arXiv preprint arXiv:1409.1556}}
  (\bibinfo{year}{2014}).
\newblock


\bibitem[\protect\citeauthoryear{Sun, Yuan, Jia, and Shum}{Sun
  et~al\mbox{.}}{2005}]%
        {sun2005image}
\bibfield{author}{\bibinfo{person}{Jian Sun}, \bibinfo{person}{Lu Yuan},
  \bibinfo{person}{Jiaya Jia}, {and} \bibinfo{person}{Heung-Yeung Shum}.}
  \bibinfo{year}{2005}\natexlab{}.
\newblock \showarticletitle{Image completion with structure propagation}. In
  \bibinfo{booktitle}{\emph{ACM Transactions on Graphics (ToG)}},
  Vol.~\bibinfo{volume}{24}. ACM, \bibinfo{pages}{861--868}.
\newblock


\bibitem[\protect\citeauthoryear{Sun, Xiao, Liu, and Wang}{Sun
  et~al\mbox{.}}{2019}]%
        {sun2019deep}
\bibfield{author}{\bibinfo{person}{Ke Sun}, \bibinfo{person}{Bin Xiao},
  \bibinfo{person}{Dong Liu}, {and} \bibinfo{person}{Jingdong Wang}.}
  \bibinfo{year}{2019}\natexlab{}.
\newblock \showarticletitle{Deep High-Resolution Representation Learning for
  Human Pose Estimation}.
\newblock \bibinfo{journal}{\emph{arXiv preprint arXiv:1902.09212}}
  (\bibinfo{year}{2019}).
\newblock


\bibitem[\protect\citeauthoryear{Ulyanov, Vedaldi, and Lempitsky}{Ulyanov
  et~al\mbox{.}}{2016}]%
        {ulyanov2016instance}
\bibfield{author}{\bibinfo{person}{Dmitry Ulyanov}, \bibinfo{person}{Andrea
  Vedaldi}, {and} \bibinfo{person}{Victor Lempitsky}.}
  \bibinfo{year}{2016}\natexlab{}.
\newblock \showarticletitle{Instance normalization: The missing ingredient for
  fast stylization}.
\newblock \bibinfo{journal}{\emph{arXiv preprint arXiv:1607.08022}}
  (\bibinfo{year}{2016}).
\newblock


\bibitem[\protect\citeauthoryear{Vo, Duong, and P{\'e}rez}{Vo
  et~al\mbox{.}}{2018}]%
        {Vo:2018:SI:3240508.3240678}
\bibfield{author}{\bibinfo{person}{Huy~V. Vo}, \bibinfo{person}{Ngoc Q.~K.
  Duong}, {and} \bibinfo{person}{Patrick P{\'e}rez}.}
  \bibinfo{year}{2018}\natexlab{}.
\newblock \showarticletitle{Structural Inpainting}. In
  \bibinfo{booktitle}{\emph{Proceedings of the 26th ACM International
  Conference on Multimedia}} \emph{(\bibinfo{series}{MM '18})}.
  \bibinfo{publisher}{ACM}, \bibinfo{address}{New York, NY, USA},
  \bibinfo{pages}{1948--1956}.
\newblock
\showISBNx{978-1-4503-5665-7}
\urldef\tempurl%
\url{https://doi.org/10.1145/3240508.3240678}
\showDOI{\tempurl}


\bibitem[\protect\citeauthoryear{Wang, Bovik, Sheikh, Simoncelli,
  et~al\mbox{.}}{Wang et~al\mbox{.}}{2004}]%
        {wang2004image}
\bibfield{author}{\bibinfo{person}{Zhou Wang}, \bibinfo{person}{Alan~C Bovik},
  \bibinfo{person}{Hamid~R Sheikh}, \bibinfo{person}{Eero~P Simoncelli},
  {et~al\mbox{.}}} \bibinfo{year}{2004}\natexlab{}.
\newblock \showarticletitle{Image quality assessment: from error visibility to
  structural similarity}.
\newblock \bibinfo{journal}{\emph{IEEE transactions on image processing}}
  \bibinfo{volume}{13}, \bibinfo{number}{4} (\bibinfo{year}{2004}),
  \bibinfo{pages}{600--612}.
\newblock


\bibitem[\protect\citeauthoryear{Xiong, Lin, Yang, Lu, Barnes, and Luo}{Xiong
  et~al\mbox{.}}{2019}]%
        {xiong2019foreground}
\bibfield{author}{\bibinfo{person}{Wei Xiong}, \bibinfo{person}{Zhe Lin},
  \bibinfo{person}{Jimei Yang}, \bibinfo{person}{Xin Lu},
  \bibinfo{person}{Connelly Barnes}, {and} \bibinfo{person}{Jiebo Luo}.}
  \bibinfo{year}{2019}\natexlab{}.
\newblock \showarticletitle{Foreground-aware Image Inpainting}.
\newblock \bibinfo{journal}{\emph{arXiv preprint arXiv:1901.05945}}
  (\bibinfo{year}{2019}).
\newblock


\bibitem[\protect\citeauthoryear{Xu and Sun}{Xu and Sun}{2010}]%
        {xu2010image}
\bibfield{author}{\bibinfo{person}{Zongben Xu} {and} \bibinfo{person}{Jian
  Sun}.} \bibinfo{year}{2010}\natexlab{}.
\newblock \showarticletitle{Image inpainting by patch propagation using patch
  sparsity}.
\newblock \bibinfo{journal}{\emph{IEEE transactions on image processing}}
  \bibinfo{volume}{19}, \bibinfo{number}{5} (\bibinfo{year}{2010}),
  \bibinfo{pages}{1153--1165}.
\newblock


\bibitem[\protect\citeauthoryear{Yeh, Chen, Yian~Lim, Schwing,
  Hasegawa-Johnson, and Do}{Yeh et~al\mbox{.}}{2017}]%
        {yeh2017semantic}
\bibfield{author}{\bibinfo{person}{Raymond~A Yeh}, \bibinfo{person}{Chen Chen},
  \bibinfo{person}{Teck Yian~Lim}, \bibinfo{person}{Alexander~G Schwing},
  \bibinfo{person}{Mark Hasegawa-Johnson}, {and} \bibinfo{person}{Minh~N Do}.}
  \bibinfo{year}{2017}\natexlab{}.
\newblock \showarticletitle{Semantic image inpainting with deep generative
  models}. In \bibinfo{booktitle}{\emph{Proceedings of the IEEE Conference on
  Computer Vision and Pattern Recognition}}. \bibinfo{pages}{5485--5493}.
\newblock


\bibitem[\protect\citeauthoryear{Yu, Lin, Yang, Shen, Lu, and Huang}{Yu
  et~al\mbox{.}}{2018}]%
        {yu2018generative}
\bibfield{author}{\bibinfo{person}{Jiahui Yu}, \bibinfo{person}{Zhe Lin},
  \bibinfo{person}{Jimei Yang}, \bibinfo{person}{Xiaohui Shen},
  \bibinfo{person}{Xin Lu}, {and} \bibinfo{person}{Thomas~S Huang}.}
  \bibinfo{year}{2018}\natexlab{}.
\newblock \showarticletitle{Generative image inpainting with contextual
  attention}. In \bibinfo{booktitle}{\emph{Proceedings of the IEEE Conference
  on Computer Vision and Pattern Recognition}}. \bibinfo{pages}{5505--5514}.
\newblock


\bibitem[\protect\citeauthoryear{Zhang, Hu, Luo, Zuo, and Wang}{Zhang
  et~al\mbox{.}}{2018}]%
        {zhang2018semantic}
\bibfield{author}{\bibinfo{person}{Haoran Zhang}, \bibinfo{person}{Zhenzhen
  Hu}, \bibinfo{person}{Changzhi Luo}, \bibinfo{person}{Wangmeng Zuo}, {and}
  \bibinfo{person}{Meng Wang}.} \bibinfo{year}{2018}\natexlab{}.
\newblock \showarticletitle{Semantic Image Inpainting with Progressive
  Generative Networks}. In \bibinfo{booktitle}{\emph{2018 ACM Multimedia
  Conference on Multimedia Conference}}. ACM, \bibinfo{pages}{1939--1947}.
\newblock


\bibitem[\protect\citeauthoryear{Zhang, Zuo, Chen, Meng, and Zhang}{Zhang
  et~al\mbox{.}}{2017}]%
        {zhang2017beyond}
\bibfield{author}{\bibinfo{person}{Kai Zhang}, \bibinfo{person}{Wangmeng Zuo},
  \bibinfo{person}{Yunjin Chen}, \bibinfo{person}{Deyu Meng}, {and}
  \bibinfo{person}{Lei Zhang}.} \bibinfo{year}{2017}\natexlab{}.
\newblock \showarticletitle{Beyond a gaussian denoiser: Residual learning of
  deep cnn for image denoising}.
\newblock \bibinfo{journal}{\emph{IEEE Transactions on Image Processing}}
  \bibinfo{volume}{26}, \bibinfo{number}{7} (\bibinfo{year}{2017}),
  \bibinfo{pages}{3142--3155}.
\newblock


\bibitem[\protect\citeauthoryear{Zheng, Cham, and Cai}{Zheng
  et~al\mbox{.}}{2019}]%
        {DBLP:journals/corr/abs-1903-04227}
\bibfield{author}{\bibinfo{person}{Chuanxia Zheng}, \bibinfo{person}{Tat{-}Jen
  Cham}, {and} \bibinfo{person}{Jianfei Cai}.} \bibinfo{year}{2019}\natexlab{}.
\newblock \showarticletitle{Pluralistic Image Completion}.
\newblock \bibinfo{journal}{\emph{CoRR}}  \bibinfo{volume}{abs/1903.04227}
  (\bibinfo{year}{2019}).
\newblock
\showeprint[arxiv]{1903.04227}
\urldef\tempurl%
\url{http://arxiv.org/abs/1903.04227}
\showURL{%
\tempurl}


\bibitem[\protect\citeauthoryear{Zhou, Lapedriza, Khosla, Oliva, and
  Torralba}{Zhou et~al\mbox{.}}{2017}]%
        {zhou2017places}
\bibfield{author}{\bibinfo{person}{Bolei Zhou}, \bibinfo{person}{Agata
  Lapedriza}, \bibinfo{person}{Aditya Khosla}, \bibinfo{person}{Aude Oliva},
  {and} \bibinfo{person}{Antonio Torralba}.} \bibinfo{year}{2017}\natexlab{}.
\newblock \showarticletitle{Places: A 10 million Image Database for Scene
  Recognition}.
\newblock \bibinfo{journal}{\emph{IEEE Transactions on Pattern Analysis and
  Machine Intelligence}} (\bibinfo{year}{2017}).
\newblock


\end{thebibliography}

\end{document}